\documentclass[11pt,twoside]{article}  
\usepackage{apn3conf}

\begin{document}   


\title{Internal Kinematics of Microstructures and Implications }
\titlemark{Kinematics of Microstructures }


\author{Luis F. Miranda }
\affil{Instituto de Astrof\'{\i}sica de Andaluc\'{\i}a (CSIC), 
       Ap. Correos 3004, 18080 Granada, Spain }


\contact{Luis F. Miranda }
\email{lfm@iaa.es }


\paindex{Miranda, L. F. }


\authormark{Miranda}


\keywords{K 3-35, IRAS 17347-3139, NGC 6884, Hu 2-1, IC 4846, kinematics  }


\begin{abstract}          
High resolution images at different wavelengths show the common presence 
of structures and microstructures in planetary nebulae (PNe), which are 
not well incorporated to the existing models for the formation of these 
objects. We summarize how studies of the internal kinematics, combined 
with the information provided by high resolution images, may help to 
establish the nature and possible origin of the observed structures as 
well as to provide information about the physical processes involved in 
the formation and evolution of PNe.

\end{abstract}


\section{Introduction}

Even though planetary nebulae (PNe) have been extensively observed, 
the processes involved in their formation continue being matter of 
debate. The lack of spherical symmetry in most PNe (e.g., Sahai \& Trauger 
1998) and the existence of jets in PNe (e.g., Guerrero et al. 2002) 
have contributed to the development of novel scenarios and models  
which provide different explanations for the main characteristics of 
PNe (see Balick \& Frank 2002 for a recent review). However, high 
resolution images show the common presence of structures and 
microstructures at subarcsecond scales, like multiple bipolar lobes, disks, 
knots, arcs (e.g., Sahai \& Trauger 1998; O'Dell et al. 2002), 
which need to be explained and incorporated to the models. Hence, 
it is necessary to obtain a ``complete'' information about 
the observed structures. In this respect, (high resolution) 
spectroscopic observations allow us to obtain the internal kinematics 
of the structures and represent an ideal complement to the information 
provided by the direct images. 

In this paper we illustrate the capabilities of the kinematic analysis 
to infer the nature of the observed structures and which role they may 
play in the formation of PNe. Particular attention will be drawn to the 
identification of jets and binary central stars from a kinematical 
analysis, and to the spatio-kinematic properties of the microstructures 
recently observed in water maser emission.

\section{Identification of Jets in PNe}

In addition to the PNe with jets already identified, high resolution 
images show that many more PNe contain structures which could be collimated 
outflows (Sahai \& Trauger 1998). Nevertheless, it has been demonstrated 
that some jet-like features observed in direct images are not related to 
collimated outflows (Goncalvez et al. 2002). In order to establish the jet 
nature of a particular feature, analysis of its kinematical properties 
is necessary. A jet should present (1) a relatively 
high radial velocity, as compared with the radial velocities observed in 
the rest of the nebula, (2) a narrow velocity width, indicating high 
collimation or, alternatively, a very large velocity width indicating 
bow-shock excitation (Solf 1994), and, in general, (3) the strong [NII] 
emission typical of jets in PNe. It should be noted that the 
radial velocity depends on the direction of the jet with respect to 
the observer, so that a low radial velocity does not necessarily excludes 
a jet nature. This is the case of the jet-like features observed 
in HST images of He\,2-90 (Sahai \& Nyman 2000), which were confirmed through 
high resolution spectroscopy to be true jets in spite of their relatively low 
radial velocity ($\simeq$ 26 km\,s$^{-1}$, Guerrero et al. 2001) which is 
due to a simple projection effect (Sahai et al. 2002). 

Spectroscopy is crucial to identify jets which cannot be observed 
in direct images because of their relatively faintness weak emission and 
projection effects. This is the case of NGC\,2392, the first PN in 
which jets have been detected (Gieseking, Becker \& Solf 1985). The jets 
in NGC\,2392 can be easily identified in long-slit spectra, at the 
appropriate position angles of the slit, as high-velocity 
($\simeq$ 190 km\,s$^{-1}$), narrow ($\simeq$ 8 km\,s$^{-1}$) and 
elongated bipolar emission features.

\begin{figure}[!ht]
\epsscale{.7}
\plotone{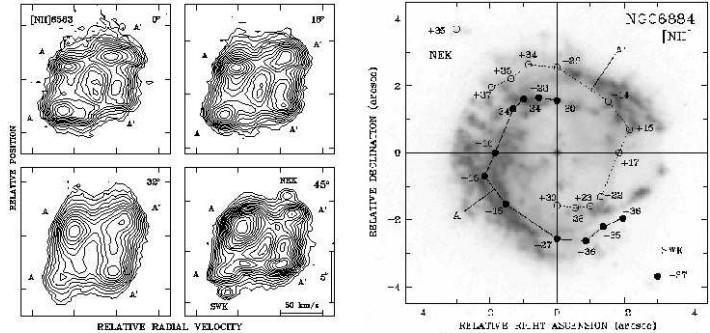}
\caption{({\it left}) Position-velocity contour maps deduced from  
high resolution long-slit spectra of the [NII] emission line from 
NGC\,6884 at four different PAs (indicated at the upper right corner). 
Kinematic components (A, A', NEK and SWK) are indicated. ({\it right}) 
Grey-scale representation of a [NII] image of NGC\,6884 obtained with the 
HST. Filled (open) circles mark the positions of the blueshifted 
(redshifted) kinematic components as deduced from a set of seven 
long-slit spectra. The numbers are radial velocities (km\,s$^{-1}$) with 
respect to the systemic velocity. (0,0) marks the position of 
the central star. } 
\end{figure}

In other PNe, structures observed in direct images can only be identified 
as jets with the aid of high resolution spectroscopy. An interesting 
case is NGC\,6884. Figure\,1 presents an [NII] image, long-slit spectra 
and the results obtained in NGC\,6884 (Miranda, Guerrero \& Torrelles 1999). 
The [NII] image shows a bright knotty structure with a very peculiar 
shape. The analysis of the kinematics shows that this structure is the 
projection of two 
narrow spirals with a high point-symmetry in space and radial velocity. The 
observed properties allow us to interpret the spirals as a precessing 
bipolar jet. Moreover, from a simple spatio-kinematic model, estimates  
for the jet velocity ($\simeq$ 55 km\,s$^{-1}$), precession 
angle ($\simeq$ 120$^{\circ}$) and precession period ($\simeq$ 
500$\times$[D(kpc)/2] yr can be deduced.

\section{Jet kinematics and binary central stars}

Binary stars are invoked in many scenarios of PN formation. Although the 
fraction of binaries among PN central stars is noticeably increasing 
(Pollaco 2004), direct detection may be difficult due to the nature, 
characteristics 
or the own evolution of the binary (e.g., Soker 1996). Binary evolution 
may have an impact in the nebular properties, which should be different from 
that of a single star. Thus, several methods have been developed in order 
to infer the possible presence of a binary central star through detailed 
analysis of the nebula. Soker (1994) and Soker, Rappaport \& 
Harpaz (1998) analyze the position of the central star with respect to the 
nebular center, as deduced from direct imaging. According to this method, 
the presence of a binary central star may be inferred from its off-centered 
location in the nebula, being possible to distinguish between wide and 
close binaries. 

On the basis of high resolution spectroscopy, Miranda et al. (2001a) 
propose that a difference between the systemic velocity of a bipolar jet 
and that of the main nebular shell may be the signature of orbital motion, 
this difference being a lower limit to the orbital velocity (see 
Miranda 2002 for details). Systemic velocity differences of $\simeq$ 10 
km\,s$^{-1}$ have been found in Hu\,2-1 and IC\,4846, two PNe with 
bipolar jets, which would imply an orbital separation $\leq$ 30 AU and 
a period $\leq$ 100 yr (Miranda 2001a,b). These orbital parameters 
suggest interacting  binaries at the center of  Hu\,2-1 and IC\,4846. 
Systemic velocity differences as low as $\simeq$ 2 km\,s$^{-1}$ 
may be detected with a spectral resolution higher than $\simeq$ 
12 km\,s$^{-1}$. Therefore, this method may be sensitive to binary central 
stars with separations up to a few tens AU.

\section{Microstructures in Water Maser Emission}

\begin{figure}[!ht]
\epsscale{.7}
\plotone{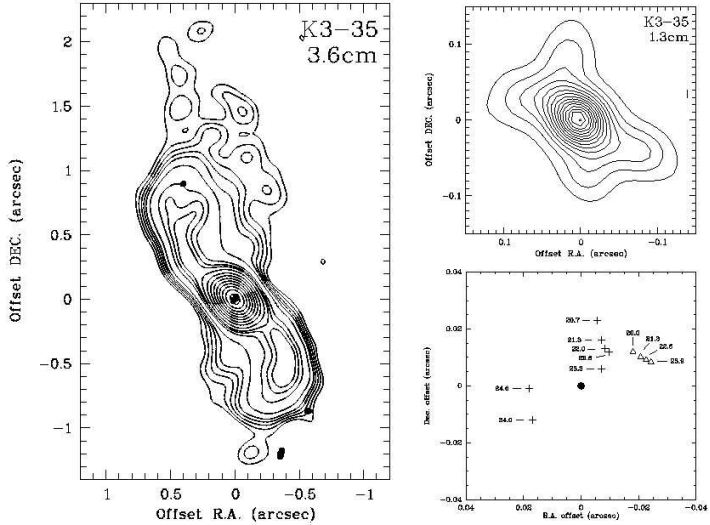}
\caption{({\it left}) Contours of the 3.6\,cm continuum emission from 
K\,3-35. Black dots represent the four regions where water maser emission 
has been detected. ({\it upper right}) Contours of the 1.3\,cm continuum 
emission. ({\it bottom right}) Position of the water maser spots at the 
center of the nebula as observed in 1999 September (crosses) and in 
2002 May (triangles). The numbers are LSR radial velocities (km\,s$^{-1}$). 
The black dot at the center marks the position of the 1.3\,cm continuum 
emission peak.}
\end{figure}

Water maser emission, typical of AGB stars (e.g., Habing 1996), persists in 
the post-AGB, proto-PN and can be detected in extremely young PNe 
(e.g., Likkel \& Morris 1988; Marvel \& Boboltz 1999; Miranda et al. 2001c). 
High resolution (VLA or VLBA) observations of water masers in 
these objects are producing exciting 
results with important implications for our understanding of PN formation. 
In the AGB or post-AGB star W43A, the water masers trace an extremely 
young, precessing bipolar jet moving at $\simeq$ 150 km\,s$^{-1}$ 
(Imai et al. 2002). In the proto-PN IRAS\,16342-3814, the water 
maser emission arises at the tips of the bipolar lobes, probably associated 
with a bow-shock, moving at $\geq$ 160 km\,s$^{-1}$ (Morris et al. 2003; 
Claussen 2004). 

The first PN in which water maser emission was detected, is K\,3-35 
(Miranda et al. 2001c). Recently, de Gregorio et al. 
(2004) have carried out a survey for water masers in a sample of 27 PNe 
and detection was obtained in IRAS\,17347-3139. Moreover, de Gregorio et al. 
(2004) deduce a T$_{eff}$ $\geq$ 26000\,K from the radio continuum emission 
for the central star of IRAS\,17347-3139, indicating an extremely young 
PN. In the following we describe the results obtained in these two PNe. 

K\,3-35 is a bipolar PN containing a bipolar jet and an extended 
equatorial disk (Aaquist \& Kwok 1989; Aaquist 1993; Miranda et al. 1998, 
2000). Figure\,2 shows VLA radio continuum maps of K\,3-35 at 3.6\,cm and 
1.3\,cm, and the location of the water masers. Water maser emission 
is detected at a radius of $\simeq$ 85 AU (for a distance of 5 kpc) from 
the center of the object and at the tips of the bipolar jet, at 
$\simeq$ 5000\,AU from the center (Miranda et al. 2001c). 

The water masers at radius $\simeq$ 85 AU probably trace the innermost 
regions of the extended equatorial disk. The disk is magnetized as indicated 
by the strong polarization of the OH maser emission from the disk 
(Miranda et al. 2001c). Changes in $\simeq$ 2.5\,yr in the positions of the 
water maser spots in the disk have been observed (see Fig.\,2; de Gregorio 
et al. 2004). Closely spaced multi-epochs 
observations are now necessary to discriminate whether the observed changes 
are due to proper motions or to a process of destruction/creation of 
water maser shells associated to an ionization/shock front. 

\begin{figure}[!ht]
\epsscale{.7}
\plotone{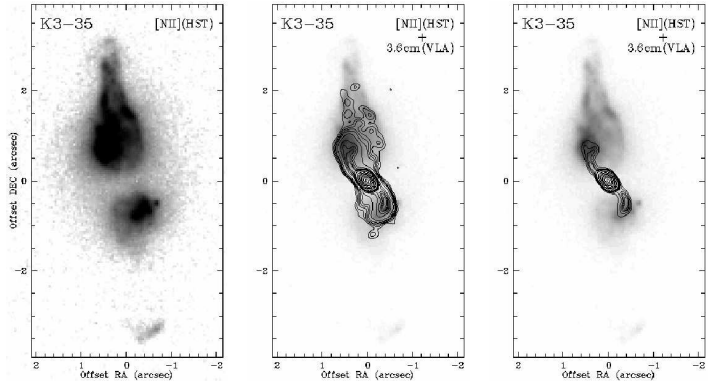}
\caption{Grey-scale maps of the [NII] image of K\,3-35 obtained with the 
HST. In the middle and right panels two sets of contours have been 
superimposed which represent the 3.6\,cm continuum emission from the 
nebula.} 
\end{figure}

The existence of the distant water masers in K\,3-35 is puzzling. Because 
the the nebula is ionized, Miranda et al. (2001c) invoke a shielding 
mechanism that prevents the distant water molecules to be destroyed 
by the stellar radiation. Given that K\,3-35 is extremely young, 
shielding could be provided by large amounts of neutral material 
in the nebula. In addition, the physical conditions required to pump the 
water maser (Marvel 1997) are not expected to exist at such enormous 
distances from the central star. The bipolar jet in K\,3-35 has been 
proposed to be the agent responsible for the excitation of the distant  
water masers, given their spatio-kinematical association 
(Miranda et al. 2001c). 
This conclusion is reinforced by comparing the 3.6\,cm continuum map 
with a HST [NII] image of K\,3-35 obtained recently by R. Sahai, which is 
shown in Figure\,3. The image shows a bipolar PNe with 
two lobes separated by a prominent dark lane (corresponding to the 
equatorial disk) and surrounded by a faint elliptical envelope. The 
previously detected bipolar knots (Miranda et al. 1998) dominate the 
emission from the lobes. The radio jet is clearly associated with these 
bright knots which may be related to a jet-envelope interaction. 
If so, the distant water masers could arise in still neutral 
clumps which are compressed and heated through this interaction.

Figure\,4 shows the HST K-band image of IRAS\,17347-3139 
(Bobrowsky, Greeley \& Meixner 1999) and the results of the VLA 
water maser observations (de Gregorio et al. 2004). The image 
shows a bipolar nebula with the main axis at PA 
$\simeq$ $-$40$^{\circ}$. The detected water maser spots trace an ellipse 
of $\simeq$ 0$\farcs$2$\times$0$\farcs$1 in size with the major axis 
oriented almost 
perpendicular to the main nebular axis. The observed morphology suggests 
that the water masers trace an equatorial disk. The 
kinematics of the water masers indicates that both rotating 
and expanding motions are present in the disk (de Gregorio et al. 2004). 

\begin{figure}[!ht]%
\epsscale{.7}
\plotone{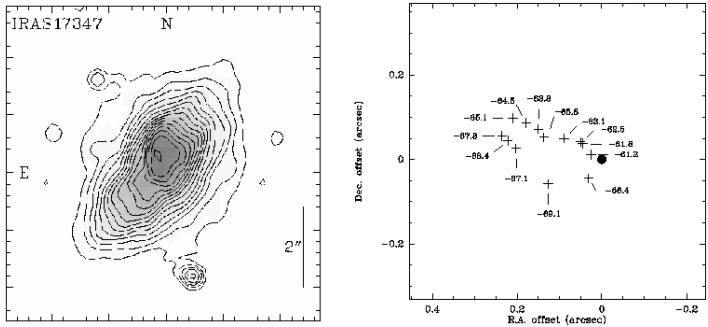}
\caption{({\it left}) Grey-scale and contour representation of the 
K-band image of IRAS\,17347-3139 obtained with the HST. ({\it right}) 
Water maser spots (crosses) observed in the nebula. The numbers are LSR 
radial velocities (km\,s$^{-1}$). Black dot at (0,0) marks the position 
of the 1.3\,cm continuum emission peak. } 
\end{figure}

The number of proto-PNe and PNe with water maser emission that 
have been observed at high resolution, is still scarce, so that 
general conclusions cannot be drawn. Nevertheless, it is interesting 
that jets are common to the few objects observed, Moreover, these jets 
operate well before the star enters its PN phase. Equatorial disks are 
also found in these objects (see references above). Remarkably, 
clear differences are observed in the kinematics of these disks. 
Expanding motions dominate in the disk of K\,3-35, as 
it is observed in more evolved PNe. The kinematics of the disk 
in IRAS\,17347-3139 is intermediate to that of the rotating disk 
detected in the proto-PN Red Rectangle (Bujarrabal et al. 2003) 
and that of expanding disks in PNe. We speculate that 
differences in disk kinematics could be related to the evolutionary 
status of the object. We note that T$_{eff}$ for the central 
star of K\,3-35 is $\geq$ 60000\,K (Miranda et al. 2000) whereas it is lower 
($\simeq$ 26000\,K) for the central star of IRAS\,17347-3139. This suggests 
that IRAS\,17347-3139 may represent an earlier stage in PN evolution than 
K\,3-35. It is possible that the fast wind in IRAS\,17347-3139 
is not yet energetic enough to swept up material in the dense equatorial 
plane and the disk preserves part of its original (rotating) motions 
before becoming a purely expanding disk typical of PNe.

\section{Conclusions}

Studies of the internal kinematics provide crucial information on  
the nature of the structures and microstructures present in PNe and 
the processes involved in the formation of these objects. The collimated 
outflow nature of jet-like features observed in direct images can be 
established by analyzing the kinematical properties of these features. 
Information about possible binary central stars can also be obtained 
from the internal kinematics of PNe with bipolar jets. Microstructures 
are also observed in recent observations of water maser emission, 
at high spatial and spectral resolution, in post-AGB stars, 
proto-PNe and extremely young PNe. These microstructures represent disks or 
jets or are associated with collimated outflows. These observations 
provide support to the idea that jets are a basic ingredient in the 
formation and evolution of some PNe and that they may be operating much 
before the star becomes a PN. The kinematics of the disks is varied and 
rotating and expanding motions can be observed as well as a combination 
of both. The kinematics of the disks could be related to the evolutionary 
stage of the object.

\vspace{0.2cm}

\noindent {\bf Acknowledgments.} I am very grateful to the organizing 
committee for the invitation to this meeting. I thank my collaborators in 
the investigation on planetary nebulae. Support from MCyT grant 
(FEDER founds) AYA2002-00376 (Spain) is also acknowledged.

%


\end{document}